\documentclass[conference,peer-reviewed]{aesconf} 

\graphicspath{{./}{figures/}}

\usepackage[utf8]{inputenc}

\usepackage{microtype}
\makeatletter
\providecommand\MT@suspend@tagging{}
\providecommand\MT@resume@tagging{}
\makeatother

\usepackage[numbers,square]{natbib}

\usepackage{booktabs}
\usepackage{color}
\usepackage{url}
\usepackage{tabularx}

\title{Evaluation of Head-Related Transfer Functions Across Five Levels of Individualisation in Virtual Reality}

\author[1]{Ludovic Pirard}
\author[1]{Katarina C. Poole}

\affil[1]{Imperial College London}

\correspondence{Ludovic Pirard}{l.pirard@imperial.ac.uk}

\lastnames{Pirard Ludovic, and Poole Katarina C.}

\shorttitle{Evaluation of HRTFs Across Five Levels of Individualisation in VR}

\begin{document}

\twocolumn[
\maketitle 

\begin{onecolabstract}
Head-related transfer functions (HRTFs) underpin spatial hearing in virtual and augmented reality systems. Whilst individual HRTFs capture listener-specific morphology, their practical limitations have led to widespread use of generic HRTFs and growing interest in synthetic approaches. Yet their relative perceptual impact remains rarely compared within a single study. In this study, we analysed data from 19 listeners that completed two virtual reality sound localisation experiments with complementary subsets of interleaved HRTF conditions enabling within-subject comparison of five conditions: individually measured, KEMAR, randomly selected non-individual measured, high-resolution scan-based synthetic and photogrammetry-based synthetic HRTFs. Test–retest stability of the individually measured baseline across sessions supported pooling across experiments and attributing differences to perceptual rather than session effects. Across HRTF conditions, lateral localisation metrics were largely insensitive to HRTF type, whereas polar-domain metrics and confusion rates showed strong HRTF dependence. Random HRTFs outperformed KEMAR on several polar metrics. High-resolution synthetic HRTFs matched individual measured performance, whilst photogrammetry-based synthetic HRTFs, alongside KEMAR, showed the greatest degradation. These findings clarify practical choices for non-individual baselines and highlight the importance of mesh resolution when using numerical synthesis for elevation-dependent localisation tasks.
\end{onecolabstract}
]

\section{Introduction}
The ability to localise sounds in three-dimensional space is fundamental to everyday listening and to perceptual immersion in virtual and augmented reality (VR/AR) applications. Spatial hearing relies on direction-dependent acoustic filtering by the listener's head, torso, and pinnae, captured by head-related transfer functions (HRTFs). These encode interaural time differences (ITDs) and interaural level differences (ILDs), which govern lateral localisation \cite{middlebrooks_sound_1991}, as well as monaural spectral cues that are essential for elevation perception and resolving front-back ambiguities \cite{langendijk_contribution_2002}. Because these cues are shaped by individual anatomy, HRTFs are inherently person-specific, and studies have consistently demonstrated that using non-individual HRTFs degrades localisation accuracy, increases confusion rates, and reduces immersion \cite{brinkmann_cross-evaluated_2019,gonzalez-toledo_spatial_2024, picinali_system--user_2023, meyer_generalization_2025}.

Despite their importance, individually measured HRTFs remain difficult to acquire at scale. Conventional measurement requires anechoic facilities, loudspeaker arrays, and precise microphone placement, making the process time-consuming, costly, and largely inaccessible outside specialist laboratories \cite{brinkmann_cross-evaluated_2019, engelSONICOMHRTFDataset2023, majdak2007multiple}. As a result, two main alternatives have emerged. The first is the use of non-individual HRTFs from existing datasets, either a fixed generic HRTF such as a KEMAR HRTF \cite{gardner_hrtf_1995} or a selected measured HRTF from another listener \cite{meyer_generalization_2025,daugintis_perceptual_evaluation_2026}. The second is numerical synthesis methods implemented in tools such as Mesh2HRTF \cite{ziegelwanger_mesh2hrtf_2015}, that simulate HRTFs directly from 3D meshes using boundary element methods. High-resolution synthetic HRTFs have been generated at scale from structured-light scans \cite{brinkmann_cross-evaluated_2019, poole_extended_2025}, while photogrammetry-based reconstruction (PR) offers a more accessible but lower-fidelity alternative \cite{pollack_perspective_2022}. This reduced geometric fidelity has been shown to degrade monaural spectral cues \cite{dellepiane_reconstructing_2008, pollack_application_2023, pollack_combination_2024}. Such differences are often quantified using numerical metrics such as spectral distortion, ITD/ILD differences, or predicted via auditory models \cite{ziegelwanger_calculation_2013, ziegelwanger_numerical_2015}. While these approaches offer scalable evaluation, they do not directly capture perceptual outcomes.


Understanding the perceptual consequences of these alternatives calls for behavioural evaluation, which typically takes the form of sound localisation \cite{middlebrooks_sound_1991, meyer_generalization_2025, middlebrooks_virtual_1999, majdak_3d_2010, daugintis_initial_2023}, quality assessment \cite{brinkmann_cross-evaluated_2019, daugintis_perceptual_evaluation_2026}, or spatial-release-from-masking tasks \cite{gonzalez-toledo_spatial_2024, daugintis_effects_2024, marggraf-turley_impact_2025, vicente_exploring_2026}.  However, such paradigms vary substantially across studies in tested conditions, protocol design, training exposure, virtual environment, response method, and stimulus configuration, making cross-study comparisons difficult. For instance, Majdak et al. (2010, 2014) varied response-pointing procedures and visual environment across conditions \cite{majdak_3d_2010, majdak_acoustic_2014}, whereas Meyer and Picinali (2025) required participants to move a cursor in front of a virtual avatar rather than directly pointing \cite{meyer_generalization_2025}, illustrating how response methods alone can differ. Frankenback et al. (2025) \cite{Frankenbach2025evaluating}, compared individual, estimated, and generic HRTFs in a dynamic binaural listening experiment, finding that individually measured and numerically synthesised HRTFs reduced vertical errors relative to KEMAR, yet even this was limited to a subset of the condition space relevant here. The broader combination of HRTFs spanning generic, randomly selected non-individual, photogrammetry-based synthetic, and high-resolution scan-based synthetic HRTFs has not been evaluated within a single unified framework, and the test-retest reliability of short VR-based localisation protocols for detecting such differences has rarely been formally characterised.

The present study addresses these gaps by combining data from two VR localisation experiments sharing the same acoustically measured HRTF baseline, spatial sampling, and validated short-form protocol, enabling within-subject comparisons across five HRTF conditions: individually measured, high-resolution scan-based synthetic, photogrammetry-based synthetic, KEMAR, and randomly selected non-individual. Behavioural localisation performance is examined, providing a comprehensive picture of how individualisation level and mesh fidelity shape spatial hearing. To our knowledge, this is the first study to compare this breadth of HRTF conditions within a unified virtual reality localisation protocol.

\subsection{Aims}
This study aims to quantify how individualisation level and acquisition method affect spatial localisation behaviour across a range of HRTF types. Three research questions are addressed:\\
\noindent\textbf{RQ1: VR Localisation test-retest reliability.}\\
Are localisation errors consistent across sessions when the same individually measured HRTFs are used, confirming the stability of the VR protocol as a basis for cross-condition comparison?\\
\noindent\textbf{RQ2: Individual versus non-individual HRTFs.}\\
Non-individual HRTFs are commonly used when individual measurements are unavailable, but do individually measured HRTFs yield meaningfully better localisation performance, and does the choice of non-individual HRTF matter?\\
\noindent\textbf{RQ3: Synthetic HRTFs versus measured HRTFs.}\\
Do HRTFs synthesised from high-resolution 3D scans or photogrammetry-reconstructed meshes yield equivalent localisation performance to individually measured HRTFs, and is there a perceptual cost to reduced mesh fidelity?

\section{Methods}


\subsection{Participants}
Twenty normal-hearing participants participated in both experiments on different days, with one excluded from further analysis due to poor correlation between re-tests, providing within-subject data for 19 participants across all conditions (mean age = 30 years, SD = 6 years, range = 22-43 years; 6 female, 13 male). All participants reported normal hearing. The study was approved by the Research Governance and Integrity Team at Imperial College London (SETREC No. 7046527) and informed consent was obtained from all participants.

\subsection{HRTF conditions}
A total of five HRTF conditions were evaluated, Table \ref{tab:HRTFconditions} summarises the conditions and their labels. Measured HRTFs were acoustically measured for all participants in accordance with the SONICOM dataset protocol \cite{engelSONICOMHRTFDataset2023}. A random HRTF condition was included as a control, similarly to \cite{pirard2026}. For this condition, HRTFs were randomly selected from the SONICOM dataset (N = 350), such that each participant was presented with a non-individual HRTF drawn from another subject. This procedure maintains the spectral and spatial characteristics of measured HRTFs (e.g. pinna and torso cues), while ensuring that the assigned HRTF does not match the listener. A KEMAR HRTF condition was included as a standard non-individual baseline. Here the HRTF was measured using a KEMAR dummy head with large pinnae and Knowles microphones, consistent with the configuration used in the SONICOM dataset protocol. This condition serves as a conventional reference for evaluating non-individual spatial filtering in binaural rendering.

\begin{table}[ht]
\centering
\begin{tabularx}{\columnwidth}{l X}
\hline
\textbf{HRTF} & \textbf{Description} \\
\hline
Measured & The participants acoustically measured HRTF\\
Random & Acoustically measured HRTF randomly selected from SONICOM dataset\\
KEMAR & Acoustically measured KEMAR HRTF with large pinnae\\
High-res scan & Synthetic HRTF generated from a high resolution scan of the participants head\\
PR scan & Synthetic HRTF generated from a photogrammetry-reconstruction of the participants head\\
\hline
\end{tabularx}
\caption{Description of the different HRTF conditions tested and their labels.}
\label{tab:HRTFconditions}
\end{table}

Two methods were used to obtain head geometry for HRTF synthesis. For the high-resolution (high-res) scan, a structured-light 3D scanner was used which projects an infrared pattern and uses stereo cameras to derive depth from its deformation to give 0.5 mm resolution (EinScan Pro 2X 2020 3D scanner). These scans were obtained as part of the SONICOM dataset \cite{engelSONICOMHRTFDataset2023} and then processed as part of the Extended SONICOM dataset \cite{poole_extended_2025}. For the PR scan, 72-image photogrammetry captures per subject were processed with Apple's Object Capture API to generate lower-resolution meshes \cite{pirard2026}. The head geometry for each of these conditions were then used to synthesise the HRTFs using Mesh2HRTF \cite{ziegelwanger_mesh2hrtf_2015}. The simulated frequency range was from 0 to 24 kHz in 150 Hz steps and acoustic propagation was solved using ML-FMM BEM solver. This was simulated across 793 source positions spanning elevation from -45$^\circ$ to 225$^\circ$ and a full 360$^\circ$ in azimuth at 5$^\circ$ intervals, matching the SONICOM measurement grid. 

All head-related impulse responses (HRIRs), sampled at 48 kHz, were windowed to a length of 256 samples using a 16-sample sine-squared fade-in and a 128-sample cosine-squared fade-out, preserving early energy while attenuating late reflections. To ensure consistent loudness across HRTF conditions, the broadband level was normalised by scaling the HRIRs such that the mean root-mean-square (RMS) level at the front (\(0^{\circ}\) elevation, \(0^{\circ}\) azimuth) between the two ears matched the KEMAR HRTF level at the same position. ITDs were extracted and removed following the threshold-based onset detection method adopted in the SONICOM dataset \cite{engelSONICOMHRTFDataset2023}.

\subsection{VR Localisation test}
\subsubsection{Apparatus}
\begin{figure}[h]
\centering
\includegraphics[width=\columnwidth]{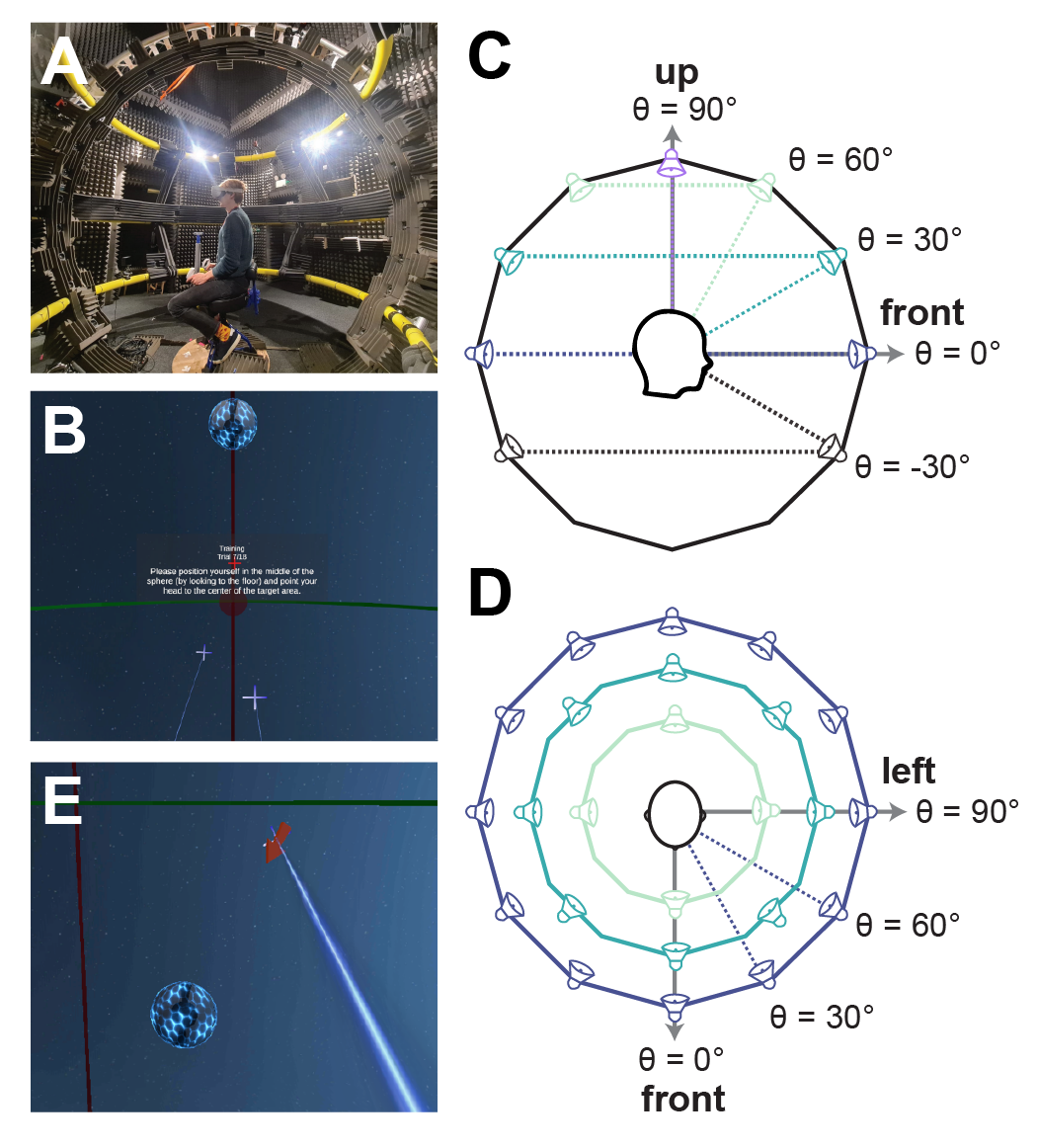}
\caption{Apparatus and source location of the virtual reality localisation test. A)  Participant seated within the loudspeaker array wearing the VR headset used for localisation testing B) Screenshot of the VR localisation task as seen by the participant, showing the virtual environment. C) Schematic of target source positions in the median plane, coloured by elevation angle. D) Top down schematic of target source positions coloured by elevation angle. E) Visual feedback display shown to the participant following each response during the training phase.}
\label{fig:Figure1}
\end{figure}

Experiments were conducted in a semi-anechoic room using a Meta Quest 2 or Quest 3 VR headset and Sennheiser HD 599SE headphones (see Fig \ref{fig:Figure1}A). The VR environment designed by Daugintis et al. \cite{daugintis_perceptual_evaluation_2026, daugintis_initial_2023}, displayed the sagittal, horizontal, and coronal planes as coloured lines on a sphere with a starry night background, providing minimal visual orientation cues (see Fig \ref{fig:Figure1}B). Participants indicated perceived sound direction with a visual laser pointer controlled via the Quest controllers, and overall playback levels averaged 65 dBA, depending on source position. The training phase presented free-field stimuli via a loudspeaker array (see Fig \ref{fig:Figure1}A; Wilmslow Audio Ltd, UK) equipped with full-range 3-inch Peerless 830987 drivers and driven by a Triad TS-PAMP8-100 amplifier.

During the main experiment, binaural audio was rendered via a standalone Unity application using the 3D TuneIn (3DTI) Toolkit \cite{3dti_2019}. The stimulus was a sequence of three Gaussian noise bursts (100 ms each), windowed with a Hann function, yielding a total duration of 300 ms. Stimuli were generated offline and convolved in real time with the selected HRTF for each trial and condition. Source positions were distributed across 33 distinct directions on the sphere, covering azimuths from \(-180^\circ\) to \(180^\circ\) and elevations from \(-30^\circ\) to \(90^\circ\), with denser sampling around the cardinal axes, following the protocol described in \cite{daugintis_perceptual_evaluation_2026, daugintis_initial_2023}(see Fig \ref{fig:Figure1}C-D). On each trial, a single virtual source was presented, and participants recorded its perceived direction by pointing and pressing the trigger. Both the three-dimensional pointing direction and the target position under each HRTF condition were logged.

\subsubsection{Procedure}
Both localisation experiments followed the same training and test procedure. Participants first completed an 87-trial training phase using a loudspeaker array to familiarise themselves with the task, response interface and externalised reference sound perception. During training, an on-screen visual target marker was initially displayed and gradually removed, with colour-coded feedback (see Fig \ref{fig:Figure1}E). In each experiment, the main test phase comprised three blocks of 99 trials, yielding 297 test trials per experiment. Across trials, the 33 source positions were presented three times each in random order with each HRTF condition interleaved. In one session, three HRTF conditions were tested: Measured, High-res scan, and KEMAR. In the other session \cite{pirard2026}, another three conditions were tested: Measured, Random, and PR scan. The data for the measured HRTF condition from the latter session was then used in all subsequent cross-condition comparisons. 20 participants completed both experiments yielding behavioural localisation data for five HRTF conditions (Measured, High-res scan, PR scan, KEMAR and Random), obtained under identical experimental protocols.

\subsection{Behavioural analysis}
Behavioural performance was quantified using standard spherical and lateral-polar metrics \cite{middlebrooks_virtual_1999, daugintis_initial_2023}. For each participant and HRTF condition, we computed the median across trials of the great-circle error, signed and absolute lateral accuracy, signed and absolute polar accuracy, polar precision, front-back confusion rate, and quadrant error rate. Front-back confusions were defined following \cite{poirier_quinot_spatial_2022} as responses falling within a \(45^\circ\) cone around the mirrored front-back position of the target. Quadrant errors were computed  following \cite{middlebrooks_virtual_1999} as a percentage of responses, restricted to trials where the lateral response was within $\pm$30$^\circ$ of the target, in which the polar error exceeded 90$^\circ$. Both front-back confusion and quadrant error rates are reported to facilitate comparison with prior literature, as studies vary in which metrics are adopted. For each metric, normality was assessed using Shapiro-Wilk tests. Where normality was satisfied, a one-way repeated-measures ANOVA was conducted with HRTF condition as the within-subject factor, followed by Tukey-HSD post-hoc comparisons. Where normality was violated, a Friedman test was used instead, followed by pairwise Wilcoxon signed-rank tests with Holm-Bonferroni correction.

\section{Results}
\subsection{RQ1: Test-retest reliability}
\begin{figure}[htbp]
\centering
\includegraphics[width=\columnwidth]{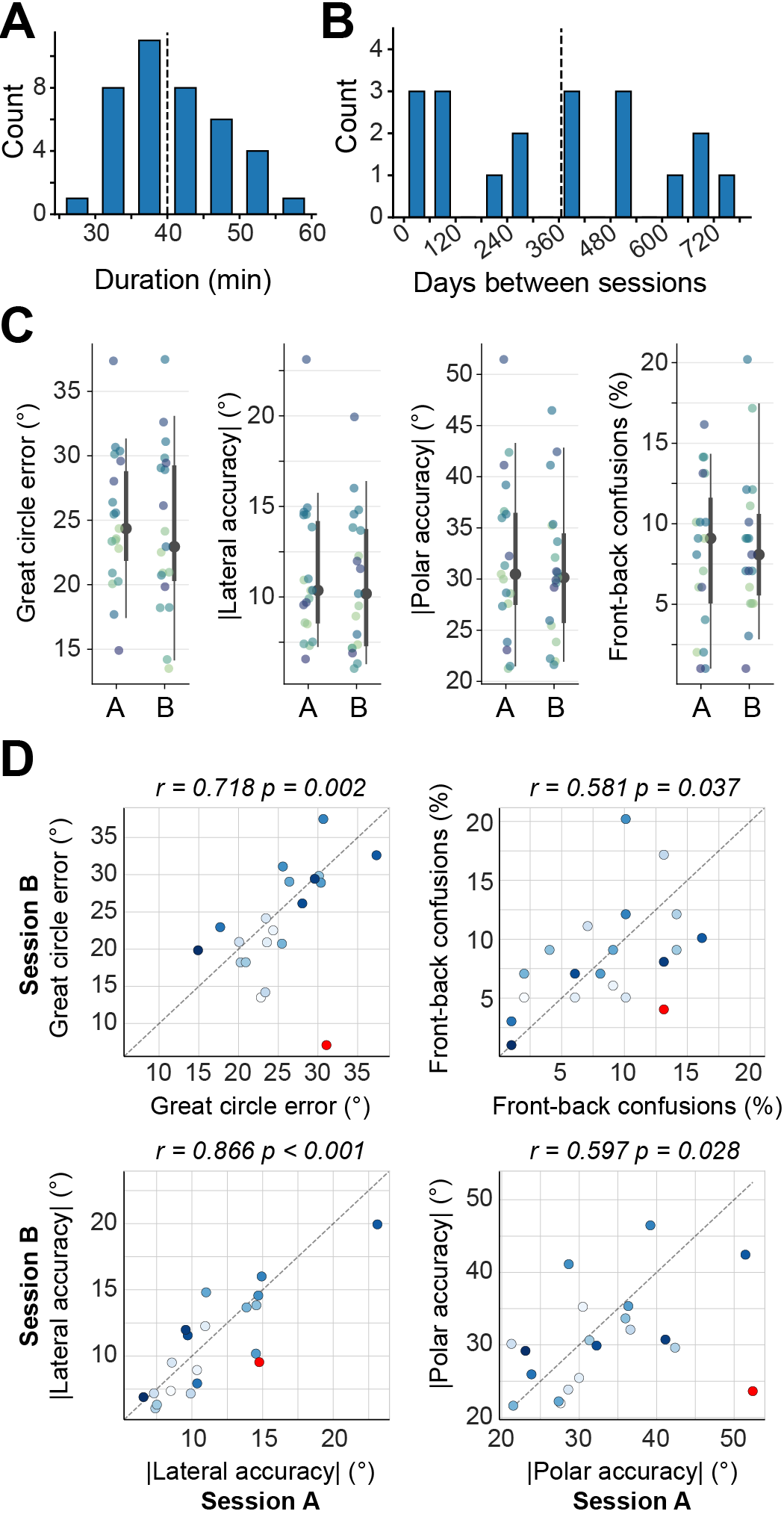}
\caption{Comparison of two sessions with the same Measured HRTF condition. A-B) Distribution of session duration (A) interval between the two experiments (B). Dashed vertical line denotes the median. C) Aggregate localisation metrics across participants. Errors bars display median, interquartile range (25th to 75th percentiles) and the 95th percentiles with each data point coloured by participant. D) Per-participant mean error metrics for Session A versus Session B. Points are coloured by participant with the outlying participant highlighted in red. Pearson correlation coefficients are reported above each panel.}
\label{fig:Figure2}
\end{figure}
First before comparing HRTF conditions, the reliability of the VR localisation protocol was established by examining consistency across two sessions (session A and session B) in which participants were presented with their individually measured HRTFs, in addition to the other HRTF conditions. Each session lasted on average 40 minutes including training (see Fig \ref{fig:Figure2}A), with approximately 10 minutes per condition. The inter-session interval varied considerably across participants (median = 364 days, range = 1 to 728 days; see Fig \ref{fig:Figure2}B), providing an ecologically varied test of protocol stability. To quantify session-to-session consistency across the key dimensions of localisation behaviour, absolute errors in the polar and lateral dimensions were calculated as well as the great circle error and front-back confusion rate for each participant and session (see Fig  \ref{fig:Figure2}C). No significant session differences were found for the great circle error ($t$(18) = 0.716, $p$ = 1; all Bonferroni corrected), lateral absolute accuracy ($t$(18) = 0.814, $p$ = 1), polar absolute accuracy ($t$(18) = 0.726, $p$ = 1), or front–back confusion rate ($t$(18) = -0.434, $p$ = 1). To further assess the individuality of localisation performance, Pearson correlations were computed between participants' scores across sessions for the measured HRTF, revealing moderate to strong relationships across all metrics: great circle error ($r$ = 0.718, $p$ = 0.002), front-back confusions ($r$ = 0.581, $p$ = 0.037), absolute lateral accuracy ($r$ = 0.866, $p$ < 0.001), and absolute polar accuracy ($r$ = 0.597, $p$ = 0.028) (see Fig  \ref{fig:Figure2}D). This revealed one outlying participant (shown in red) who demonstrated poor test-retest reliability and was therefore excluded from all analyses. Collectively, these findings confirm the systematic stability of the VR localisation protocol across an inter-session interval of approximately one year, and validate pooling data across experiments.

\subsection{Effect of HRTF Condition on Localisation}
\begin{figure*}[htbp]
    \centering
    \includegraphics[width=\textwidth]{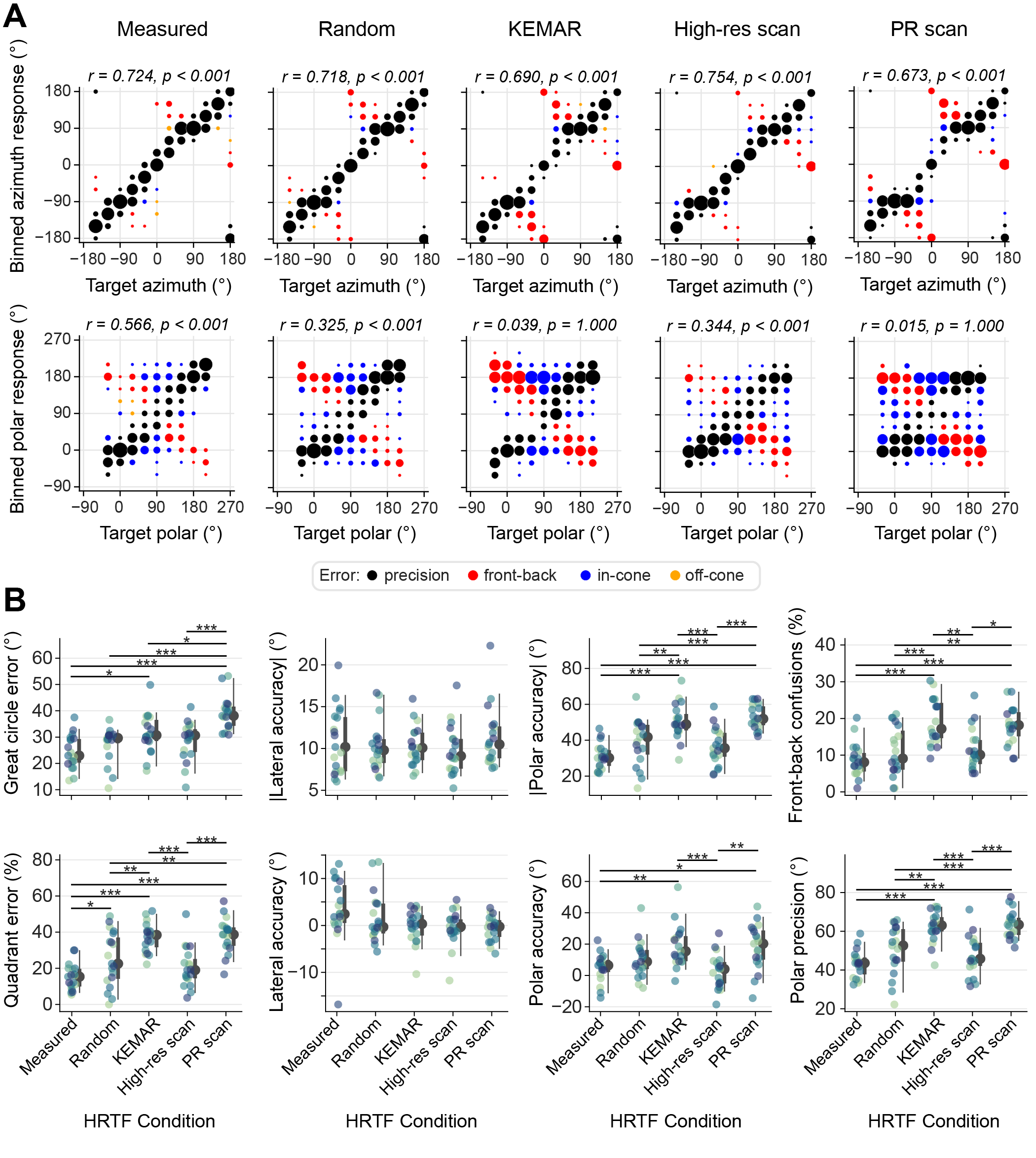}
    \caption{Behavioural localisation performance across the five HRTF conditions (N = 19). A) Binned localisation plots with correlation coefficients computed for both the azimuth and polar planes and scatter points are coloured by error type. B) Aggregate localisation metrics across participants for each HRTF condition. Errors bars display median, interquartile range (25th to 75th percentiles) and the 95th percentiles with each data point coloured by participant. Bars shows significant post-hoc pairwise comparisons; * = $p$ < 0.05, ** = $p$ < 0.01, *** = $p$ < 0.001.}
    \label{fig:Figure3}
\end{figure*}
Localisation performance across the five HRTF conditions was assessed using eight metrics capturing complementary aspects of spatial hearing: great circle error, lateral signed and absolute accuracy, polar signed and absolute accuracy and precision, front-back confusion and quadrant error rates (see Fig \ref{fig:Figure3}). Significant main effects of HRTF condition were found for the great circle error ($\chi^2$(4) = 34.65, $p$ < 0.001), polar absolute accuracy ($F$(4, 72) = 27.76, $p$ < 0.001), front-back confusion ($\chi^2$(4) = 41.11, $p$ < 0.001), quadrant error ($F$(4, 72) = 24.91, $p$ < 0.001), signed polar accuracy ($F$(4, 72) = 12.47, $p$ < .001), and polar precision ($F$(4, 72) = 28.48, $p$ < 0.001), but not for lateral absolute or signed accuracy (both $p$ > 0.05), confirming that the HRTF condition affected polar but not lateral localisation.

\subsubsection{RQ2: Effect of HRTF individualisation}
To examine the effect of HRTF individualisation, post-hoc comparisons focused on differences between the individually measured HRTF and the two non-individual conditions: the KEMAR and a randomly selected measured HRTF. Horizontal plane target-response correlations were high across all conditions (measured: $r$ = 0.724, random: $r$ = 0.718, KEMAR: $r$ = 0.690; see Fig \ref{fig:Figure3}A), consistent with the absence of lateral main effects. The median plane revealed a striking dissociation where the measured HRTF yielded a moderate elevation correlation ($r$ = 0.566, $p$ < 0.001), while KEMAR showed near-zero elevation localisation ($r$ = 0.039, p = 1) and the random HRTF an intermediate value ($r$ = 0.325, $p$ < 0.001).

This pattern was consistent across polar localisation metrics (see Fig \ref{fig:Figure3}B). Polar absolute accuracy was significantly worse for KEMAR (48.8$^\circ$) relative to measured (30.1$^\circ$, $p$ < 0.001), whereas the random HRTF (41.8$^\circ$) did not differ significantly from measured ($p$ = 0.078). KEMAR also showed significantly worse polar precision than both measured and random (all $p$ < 0.001), and a systematic bias towards rear responses relative to measured ($p$ = 0.007), while the random HRTF showed no such bias ($p$ = 0.639). Front-back confusion rates were markedly elevated for KEMAR (17.2\%), relative to measured (8.1\%) and random (9.1\%, both $p$ < 0.001), with random performing comparably to measured. Notably, despite this pattern of broad equivalence between random and measured HRTFs, quadrant errors were significantly higher for the random HRTF than measured (22.6\% vs 15.4\%, $p$ = 0.039), suggesting that randomly selected HRTFs introduce polar localisation errors beyond clean front-back confusions. Overall, non-individual HRTFs primarily impaired polar localisation, with the KEMAR producing the most severe degradation and randomly selected HRTFs offering only partial preservation of spatial cues.

\subsubsection{RQ3: Effect of mesh resolution on synthetic HRTF performance}

To examine the effect of mesh resolution on synthetic HRTF performance, post-hoc comparisons focused on the two scan-based synthetic conditions, high-resolution and PR, relative to the individually measured HRTF and the non-individual baselines established in RQ2. In the horizontal plane, the high-resolution synthetic HRTF produced the strongest azimuth correlation of all conditions ($r$ = 0.754), while the PR synthetic HRTF showed the weakest ($r$ = 0.673, see Fig \ref{fig:Figure3}A). In the median plane, the high-resolution synthetic HRTF yielded a moderate elevation correlation ($r$ = 0.325, $p$ < 0.001), well above KEMAR ($r$ = 0.039) but below measured ($r$ = 0.566). The PR synthetic HRTF showed near-zero elevation tracking ($r$ = 0.015), comparable to the KEMAR HRTF condition.

Across all polar metrics, the high-resolution synthetic HRTF performed comparably to the individually measured HRTF: polar absolute accuracy (35.6$^\circ$ vs 30.1$^\circ$), polar precision (45.8$^\circ$ vs 43.6$^\circ$), front-back confusion (10.1\% vs 8.1\%), and quadrant error (19.0\% vs 15.4\%, all $p$ = n.s.), and significantly outperformed KEMAR on polar accuracy, precision, and front-back confusion (all $p$ < 0.003, see Fig \ref{fig:Figure3}B). By contrast, the PR synthetic HRTF was  significantly worse than measured across all metrics (all $p$ < 0.011) and comparable to KEMAR in polar accuracy, front-back confusion, and signed polar bias. Notably, the PR scan HRTF also performed significantly worse than the randomly selected non-individual HRTF for great circle error, polar accuracy, front-back confusion, and quadrant error (all $p$ < 0.010), suggesting that photogrammetry-based synthesis in its current form offers no perceptual advantage over a non-individual HRTF, whereas high-resolution scan-based synthesis yields localisation performance indistinguishable from individual acoustic measurement.

\section{Discussion}
\subsection{Reliability and utility of the VR localisation protocol}
The heterogeneity of behavioural paradigms for HRTF evaluation in sound localisation tasks complicates cross-study comparison and raises questions about which protocols yield stable, interpretable condition differences \cite{meyer_generalization_2025, majdak_3d_2010, majdak_acoustic_2014, Frankenbach2025evaluating}. In the present work, the same short VR localisation protocol \cite{daugintis_perceptual_evaluation_2026, daugintis_initial_2023, pirard2026} ($\approx$ 10 min per condition; $\approx$ 40 min total including training) was used across both sessions with complementary subsets of interleaved HRTF conditions, avoiding a blocked design in which prolonged exposure to a single HRTF may promote perceptual adaptation and obscure true condition differences. Whilst the training phase familiarised participants with externalised free-field sound perception, the degree to which binaural rendering was perceived as externalised was not formally assessed. The significant target–response elevation correlations for the measured HRTF suggest structured spatial judgements, though in-head localisation cannot be fully excluded. This test-retest stability has two important implications. First, it justifies directly comparing HRTF conditions across the two experiments, indicating that observed differences are genuine perceptual effects rather than session-to-session variability. Second, it supports short VR-based localisation protocols as reliable outcome measures for HRTF evaluation more broadly, which is non-trivial given the methodological heterogeneity reported in the literature. 



\subsection{Random non-individual HRTFs as a practical baseline beyond KEMAR}
The conventional use of a single generic KEMAR HRTF as the non-individual baseline in previous studies on sound localisation, speech release from masking and spatial audio quality assessment warrants reconsideration. Although KEMAR and randomly selected measured HRTFs showed similar performance in some metrics, the random HRTF showed enhanced performance in polar localisation and confusion rates. A plausible explanation is that randomly selected measured HRTFs preserve naturally occurring human acoustic variability in pinna, head, and torso filtering, thereby retaining realistic (albeit non-individual) spectral structure that a fixed mannequin representation cannot fully capture. In this sense, randomly selected measured HRTFs provide a more ecologically valid approximation of non-individual human spatial filtering than a single generic mannequin HRTF, and therefore a fairer baseline for evaluating non-individual rendering strategies. Moreover, because KEMAR reflects the acoustic properties of a single fixed anatomy, its repeated use as a universal non-individual baseline may introduce consistent directional biases across studies that a more variable human-derived HRTF would not, a possibility that warrants further investigation.

These findings suggest that studies relying solely on a KEMAR HRTF may underestimate the performance achievable without individualisation, and future work should consider including a randomly selected measured HRTF as an additional baseline. For VR and AR applications where individual measurement is infeasible, selecting a random HRTF from an open dataset such as SONICOM requires no greater effort than using a KEMAR HRTF but may yield meaningfully enhanced elevation perception and fewer front-back confusions.


\subsection{Importance of levels of morphology detail in synthetic HRTFs for sound localisation}
Different levels of mesh resolution have previously been evaluated using numerical metrics in Ziegelwanger et al. (2013)  \cite{ziegelwanger_calculation_2013}, and subsequently with the Baumgartner2014 auditory model and a sound localisation task (N = 3) in Ziegelwanger et al. (2015) \cite{ziegelwanger_numerical_2015}. However, synthetic HRTFs derived from high-resolution scan and photogrammetry reconstructed meshes have not been evaluated side by side and against individually measured HRTFs in a sound localisation test with a large participant sample. Our behavioural data reveal clearly different outcomes for both synthetic conditions. The performance of PR scan synthetic HRTFs, although derived from individual meshes, indicates that at this current reconstruction fidelity and mesh resolution, they do not preserve sufficient pinna detail to encode the monaural spectral cues required for robust elevation localisation. In contrast, high-resolution synthetic HRTFs were not significantly different from the individual measured HRTF across the evaluated metrics. 

Together, these findings highlight the importance of high-resolution and fidelity in head-and-pinna geometry when computing synthetic HRTFs with Mesh2HRTF. Furthermore, the equivalence of high-resolution synthetic and individually measured HRTFs has important practical implications. Where acoustic measurement is unavailable, high-resolution scan-based synthesis represents a viable alternative that preserves the perceptual benefits of individualisation and is considerably more scalable than anechoic measurement, supporting its use for individualised spatial audio in VR and AR applications.

\section{Conclusion}
By merging two VR localisation studies with matched individual baselines and protocols, we compared five HRTF conditions within a unified behavioural framework. Lateral metrics were insensitive to condition, whereas elevation-related measures and confusion rates revealed clear degradation for photogrammetry-based synthetic and KEMAR HRTFs relative to individually measured and high-resolution synthetic conditions. Randomly selected non-individual HRTFs performed comparably to or better than KEMAR across several metrics. The test-retest stability of the VR localisation protocol justifies cross-experiment comparisons as reflecting genuine perceptual effects and supports short VR-based protocols as reliable outcome measures for HRTF evaluation


Future work will focus on correlating the numerical metrics and auditory model predictions obtained in Pirard et al. (2026) \cite{pirard2026} with the behavioural data acquired here. Notably, \cite{pirard2026} examines numerical and auditory model analyses across a substantially larger sample (N = 150), and establishing how these relate to behavioural localisation outcomes represents a key next step. The breadth of HRTF types evaluated within a single controlled protocol may offer particular opportunities for identifying which computational measures best predict perceptual differences across conditions, ultimately contributing to a deeper understanding of the acoustic and anatomical factors that shape individual spatial hearing and informing the design of more accurate spatial audio rendering systems for VR and AR.


\section{Acknowledgements}
This work was supported by the European Union within the project SONICOM (Grant No. 101017743, RIA action of Horizon 2020) and the CherISH European Doctorate Network (Grant No. 101120054, HORIZON-MSCA-2022-DN-01, Marie Skłodowska-Curie Actions, Horizon Europe).

\bibliographystyle{jaes}

\bibliography{refs}

\end{document}